\documentstyle[12pt,aps,prc,epsfig,preprint]{revtex}
\tightenlines
\begin{document}

\begin{titlepage}
\title{\vspace*{10mm}\bf $K^-$ and $\bar p$
deeply bound atomic states}
%\vspace{6pt}

\author{ E.~Friedman and A.~Gal \\
{\it Racah Institute of Physics, The Hebrew University, Jerusalem 91904,
Israel\\}}

%\vspace{4pt}
\maketitle
\begin{abstract}

The strongly absorptive optical potentials 
$V_{{\rm opt}}$ which have been deduced
from the strong-interaction level shifts and widths in X-ray spectra
of $K^-$ and $\bar p$ atoms produce effective repulsion leading
to substantial suppression of the {\it atomic} wave functions within
the nucleus. The width of atomic levels then saturates as function
of the strength of Im $V_{{\rm opt}}$. We find that `deeply bound' atomic
states, which are inaccessible in the atomic cascade process, 
are generally narrow, due to this mechanism,
over the entire periodic table
%reaching values of $\Gamma_{K^-} \sim 1.7$ MeV and
%$\Gamma_{\bar p} \sim 1.9$ MeV for the $1s$ state in Pb, 
and should be reasonably well resolved. These predictions are insensitive
to $V_{{\rm opt}}$, provided it was fitted to the observed X-ray spectra.
In contrast, the {\it nuclear} states bound by $V_{{\rm opt}}$ are very
broad and their spectrum depends sensitively on details of $V_{{\rm opt}}$.
We discuss production reactions  for $K^-$ atomic states using
slow $K^-$ mesons from the decay of the $\phi$(1020) vector meson,
and the ($\bar p, p$) reaction for $\bar p$ atomic states.
Rough cross-section estimates are given.
\newline \newline
$PACS$: 21.65.+f; 36.10.Gv
\newline
 \newline\newline
%\vspace{1cm}
Corresponding author: E. Friedman,\newline
Tel: +972 2 658 4667,
Fax: +972 2 658 6347, \newline
E mail: elifried@vms.huji.ac.il
\end{abstract}
\centerline{\today}
\end{titlepage}

\section{Introduction}
\label{sec:int}

Deeply bound hadronic atom levels have been observed recently
for pions \cite{Yam96,Yam98},
using the ($d,^3$He) recoilless reaction \cite{HTY91}
on $^{208}$Pb, following earlier predictions that the $1s$ and $2p$
atomic levels in pionic Pb have widths of order 0.5 MeV, significantly
less than the approximately 1.5 MeV spacing \cite{FS85,TYa88}.
This striking narrowness is due to the well established repulsive $s$-wave
part of the pion-nucleus potential at threshold which pushes the
corresponding atomic wave functions out of the nucleus such that their
overlap with the nucleus, and hence with the imaginary part of the potential,
is substantially reduced. A similar, but not as favourable situation
might occur for $\Sigma^-$ atoms due to the inner repulsion of the
$\Sigma$ nucleus potential \cite{BFG94a,BFG94b,MFGJ95}. 
Such deeply bound hadronic atom levels are inaccessible via the
atomic cascade process because of the large absorptive widths compared
to the radiation widths. The interest
in observing such `deeply bound' atomic levels stems from anticipating
a larger overlap of the corresponding wave functions with the nuclear
density profile, and hence a greater sensitivity to the hadron-nucleus
strong interaction than is the case for normal states.

The other hadronic atom
species which have been studied experimentally, consisting of $K^-$ and
$\bar p$ atoms, do not appear at first sight likely candidates for
narrow deeply bound states, since the real part of the hadron-nucleus
potential at threshold is known for these hadronic species to be
strongly attractive and, furthermore, the imaginary (absorptive) part
is particularly strong, reaching (absolute) values of order 50 - 100
MeV inside nuclei \cite{BFG97}.
 However, it was noted years ago by Krell \cite{Kre71}
and by Koch  et al. \cite{KSW71,KSW72} for
$K^-$ atoms, and by Green et al. \cite{GW82,GSW83}
for $\bar p$ atoms, that
the strong absorptive potential  for these species
made the optical potential at threshold effectively
repulsive, even in the presence of a relatively strong
attractive real potential, as evidenced by the repulsive
atomic level shifts reproduced by the fitted optical potential 
$V_{{\rm opt}}$.
It was argued in these works that for strong
absorptivities the atomic wave functions $\psi$ are substantially
suppressed within the nucleus and that the atomic level
widths might saturate as function of Im $V_{{\rm opt}}$.
However, the latter effect has not been studied
systematicaly and quantitatively for {\it realistic} situations.
In particular, there have been no reliable calculations
for the extent of saturation for widths of atomic levels,
resulting from the suppression of
$\left| \psi \right|^2$ as function of the strength
of the absorptive potential.

Wave function damping and subsequent
saturation effects due to absorptive potentials 
are known in elastic scattering.
The introduction of absorptivity, in terms of an imaginary part $W$
of the optical potential $V_{{\rm opt}} = U - {\rm i}W$, often induces repulsion
into the description of elastic scattering, leading to expulsion of
the elastic channel wave function from the nuclear interior.
For example, in the eikonal description of nuclear reactions
\cite{Gla59}, the wave function is exponentially attenuated by the
factor

\begin{equation}
\label{equ:eik}
\left| {\psi \left( {b,z} \right)} \right|^2=\exp \left( {-{2 \over
{\hbar v}}\int\limits_{-\infty }^z {W \left( {b,z'} \right)dz'}}
\right)
\end{equation}
where $v$ is the hadronic projectile velocity
and ${\bf r}$ = (${\bf b},z$).
A consequence of this attenuation is that the total reaction
cross section $\sigma_R$, which measures the loss of incident flux
into the absorptive channels, saturates as function of the strength
of $W$, so that

\begin{equation}
\label{equ:sigR}
\sigma _R=\int {\left( {1-\left| {\psi \left(
{b,z\to \infty } \right)} \right|^2} \right)d^2b\to \pi R^2}
\end{equation}
for square-well potentials of radius $R$, as $W$ is made sufficiently strong.

A similar situation must occur also for hadronic atoms where the
`outer' atomic levels (analogous to the elastic channel above)
are coupled to the `inner' absorptive channels by the nuclear
interaction. This coupling is particularly effective in $K^-$
and in $\bar p$ atoms due to the strong one-nucleon
absorption processes
%\begin{equation}
%\label{equ:abs}
$K^{-} + N \rightarrow \pi + Y \;\;\; ,
\;\;\; \bar p + N \rightarrow~{\rm mesons}$,
%\end{equation}
respectively.
In fact, for a Schr\"odinger-type equation, the width $\Gamma$ of the level is
given {\it exactly} (i.e. not perturbatively) by:

\begin{equation} \label{equ:gamma}
\frac{\Gamma}{2}= \frac{\int W(r) | \psi({\bf r}) | ^2  d {\bf r}}
{\int | \psi({\bf r}) | ^2  d {\bf r}}\quad,
\end{equation}
which is the straightforward extension of the equality (\ref{equ:sigR}) into
the bound state regime. Clearly if $|\psi |^2$ is expelled from
the nuclear domain where $W=-$Im~$V_{{\rm opt}}$ is operative, then the width
$\Gamma$ may also saturate.

In a recent Letter \cite{FGa99} we demonstrated that the width
of $K^-$ {\it atomic} levels generally saturates, and that
consequently $K^-$ deeply bound atomic states which are
inaccessible in the atomic cascade process are by far narrower
than a simple extrapolation from the observed states which are
accessible in the atomic cascade process would lead to.
Here we report on a comprehensive study of this saturation
phenomenon for widths in $K^-$ atoms, and in parallel also for
$\bar p$ atoms. We find that deeply bound atomic
states are relatively narrow, becoming as little broad as
$\Gamma_{K^-} \sim 1.7$ MeV and
$\Gamma_{\bar p} \sim 1.9$ MeV for the $1s$ state in Pb,
and could in many instances be sufficiently well resolved.
Our predictions for these states are insensitive to within
a few percents to details of $V_{{\rm opt}}$, provided it was fitted to the
observed X-ray spectra. In contrast, the {\it nuclear} states bound
essentially by $V_{{\rm opt}}$ are very broad and their spectrum depends
sensitively on details of $V_{{\rm opt}}$. Lastly, we discuss a few candidate
production reactions: ($K^-, p$) 
and $(\phi,K^+)$ for $K^-$ deeply bound atomic states,
using slow $K^-$ mesons from the decay of the $\phi$(1020) vector meson
or from its secondary interactions,
and ($\bar p, p$) for $\bar p$ deeply bound atomic states.
We give production cross-section estimates which are new for the
($K^-, p$) and $(\phi,K^+)$ reactions, and which for ($\bar p, p$) follow
the calculations already reported in Refs. \cite{HJM83,GK84,BDS85}.
The latter reaction was used at LEAR \cite{Gar85} searching,
with no success, for deeply bound {\it nuclear} states.

\section{Methodology}\label{sec:meth}

The interaction of hadrons at threshold with the nucleus
is described  in this work,
as well as in all our past work on hadronic atoms, by the Klein-Gordon (KG)
 equation of the form:

\begin{equation}\label{equ:KG1}
\left[ \nabla^2  - 2{\mu}(B+V_{{\rm opt}} + V_c) + (V_c+B)^2\right] \psi = 0~~ ~~
(\hbar = c = 1)
\end{equation}
where $\mu$ is the hadron-nucleus reduced mass,
$B$ is the complex binding energy
 and $V_c$ is the finite-size
Coulomb interaction of the hadron with the nucleus, including 
vacuum-polarization terms.
Equation (\ref{equ:KG1}) assumes that $V_{{\rm opt}}$ behaves as a Lorentz scalar.
If $V_{{\rm opt}}$ is assumed to behave as a Lorentz time-component
of a four vector, then additional terms involving $V_{{\rm opt}}$, of
order ($V_{\rm c}$+Re~$B)/\mu$ and $\Gamma/2\mu$, arise.
We have  verified
that this neglect is well justified for atomic states.
The use of the KG equation for $\bar p$ atoms is also justified as long as 
spin effects are negligible and one is interested in ($2j+1$) averaging.

For spherically symmetric potentials, the bound state solutions of the
KG equation (\ref{equ:KG1}) are of the usual form
$\psi_{nlm}({\bf r})=(u_{nl}(r)/r)Y_{lm}(\theta,\phi)$.
Since $V_{{\rm opt}}$ is complex, the radial wave functions are also complex
and satisfy, for each value of the angular momentum $l$, the following
generalized orthogonality relationship:

\begin{equation}
\label{equ:ortho}
\int\limits_0^\infty [1-(V_{\rm c}+B_{n,n'})/\mu]u_{n,l}(r)u_{n',l}(r)
dr=0,~~(n \ne n')
\end{equation}
where $B_{n,n'}=(B_n+B_n')/2$. For $n=n'$, the normalization integral
on the left-hand side remains arbitrary. 
It could be fixed by replacing the zero
on the right-hand side of Eq. (\ref{equ:ortho}) by $\delta _{n n'}$. We chose not
to  follow this prescription, but rather kept up with the Schr\"odinger
`decaying state' normalization \cite{GCP76}

\begin{equation}
\label{equ:norm}
\int\limits_0^\infty u^2_{nl}(r) dr = 1.
\end{equation}
Here, two conditions are incorporated: (i) a normalization condition
on (Re $u$)$^2-$(Im $u$)$^2$, and (ii) the orthogonality of Re $u$ and
Im $u$ which usually is satisfied by Im $u$ having one node more than
the number of `nuclear' nodes possessed by Re $u$. This enables to 
classify bound states by the node structure of Re $u$ as is outlined below.

Another consequence of the KG equation (\ref{equ:KG1}) is the following
exact integral expression for the width $\Gamma = 2~$Im $B$:

\begin{equation} \label{equ:gamma2}
\frac{\Gamma}{2}= \frac{\int W(r) | \psi({\bf r}) | ^2  d \bf {r}}
{\int | \psi({\bf r}) | ^2 [1-({\rm Re} B+V_c)/\mu]  d \bf {r}}\quad,
\end{equation}
generalizing Eq. (\ref{equ:gamma}) 
 due to the presence of the quadratic $(V_{\rm c}+B)^2$ term
in Eq. (\ref{equ:KG1}).

\subsection{Optical potentials}

Several optical potentials were used in order to 
check the model dependence
of the results.  In all cases the potentials 
were obtained from fits to experimental data for normal hadronic atoms.
The basic form of the potential is the 
phenomenological density dependent (DD) potential of Friedman
 et al. \cite{FGB93,FGB94}, given by:

\begin{equation}\label{equ:DD}
2\mu V_{{\rm opt}}(r) =
 -{4\pi}(1+{\frac{\mu}{m}})b(\rho)
\rho(r) \;\;\; ,
\end{equation}

\begin{equation}\label{equ:b}
b(\rho)= b_0+B_0(\frac{\rho(r)}{\rho_0})^\alpha \;\;\; ,
\end{equation}
\noindent

\noindent
where $b_0$ and $B_0$ are complex parameters determined from fits to
the data, $m$ is the mass of the nucleon, $\rho(r)$
is the nucleon-center  density distribution normalized to the number of nucleons
$A$ and $\rho_0=$0.16 fm$^{-3}$ is a typical central nuclear density.
For $B_0=0$, the usual `$t\rho$' potential is obtained. For
kaonic atoms we  used either a `$t\rho$' potential with the values
$b_0$=0.62+i0.92 fm,  or the DD potential with
$b_0$=$-$0.15+i0.62 fm, $B_0$=1.66$-$i0.04 fm, $\alpha=0.24$, with the
latter potential providing an improved fit to the data and also respecting
the low density limit
(c.f. Table 6 of Ref.\cite{BFG97}). 
It is interesing to note that these potentials are
very different from each other in the nuclear interior, e.g. the real 
attractive potential is about    
80 MeV deep for the `$t\rho$' model and 180 MeV deep for the DD model.
The nuclear densities used for kaonic atoms were mostly of the `macroscopic'
(MAC) type, obtained from phenomenological fits to electron scattering
data by unfolding the proton charge distribution. 
Some calculations were repeated with `single particle'
(SP) nuclear densities \cite{BFG97}.

For antiprotons two forms of potential were used, a `$t\rho$'
form  as discussed above or its $p$-wave extension
given by:
\begin{equation} \label{equ:EE1}
2\mu V_{{\rm opt}}(r) = q(r) + {\bf \nabla} \cdot \alpha(r) {\bf \nabla}
\end{equation}

\noindent
with

\begin{equation} \label{equ:EE1s}
q(r)  =  -4\pi(1+\frac{\mu}{m})b_0\rho(r)
\end{equation}

\begin{equation} \label{equ:EE1p}
\alpha (r)  =  4\pi(1+\frac{\mu}{m})^{-1}c_0\rho(r).
\end{equation}

\noindent
The nuclear densities were always of the SP type, 
as these lead to better
fits \cite{BFG95} 
to the data, presumably because they provide a more realistic 
description  than the MAC densities
at large radii 
on account of having slopes determined by the least bound nucleons.
The parameter values were 
(c.f. Table 8 of Ref.\cite{BFG97}) $b_0~=~2.5+$i$3.4$ fm
for the $t\rho$ potential, and $b_0=4.5+$i$4.5$ fm, $c_0=-4.0-$i$2.4$ fm$^3$
for the $p$-wave extension.

%\section{Results} \label{sec:results}
\subsection{Numerical procedure} \label{sec:numer}

Hadronic atoms are physical systems where some unusual features 
play important role. Whilst effects of the strong interaction are the
main topic of studies of such atoms, in most cases
the system is dominated by the Coulomb interaction between the hadron
and the nucleus. 
 The strong interaction is usually confined
(at least for `normal' hadronic atom states)  to a small (nuclear) region
of the atom but its strength is such that perturbation approach is totally
inadequate. The wave function is modified very strongly in the nuclear
region, causing long range effects throughout the atom.
For this reason care must be exercised in the numerical 
description of such systems.

The first obvious test is to make sure that the  binding energy obtained
numerically  for the point charge Coulomb interaction 
 agrees with the analytic 
KG binding energy (except that the test is not possible for $s$-states
in nuclei with $Z>$137/2). This is easily fulfilled to an accuracy of 5-7
significant digits. 
Note that, except for the tests, we always use the Coulomb
potential due to the extended nuclear charge distribution and include 
also the first order vacuum polarization potential.

Another test which is particularly relevant in the
present context relates to the width of the level. The width is given by
 twice the imaginary part of the complex binding energy $B$ as obtained
from solving numerically the KG 
equation (\ref{equ:KG1}). 
The width is also given  exactly by 
the integral expression Eq. (\ref{equ:gamma2}).
We routinely get agreement to at least 5 digits between the widths calculated
for realistic potentials 
using the two methods. 
The approximate expression for the width given by  
Eq. (\ref{equ:gamma}) misses the exact result  by 0.2 to 0.5\%
for atomic states.

Last but not least is the need to ensure that the calculated atomic states
are correctly classified.  
As is discussed later, in the
present case of strongly attractive potentials there are 
{\it nuclear} bound states in addition to the atomic states, and there
are subtle connections between the two kinds of wave functions that greatly
affect the latter.
In all cases we have also calculated the energies of strongly
bound nuclear states whose widths are typically tens of MeV up to over
100 MeV. As such they are insignificant  experimentally
but their existence affect markedly the atomic states, as the
wave functions of the latter satisfy the generalized orthogonality  
relationship Eq. (\ref{equ:ortho}) with the wave functions of the
former, for each value of the angular momentum $l$. This requirement together
with the dependence of the calculated widths on the strength of the
imaginary potential help to identify states unambiguously as either
deeply bound atomic states or nuclear states.

\section{Energy spectra} \label{sec:spectra}

\subsection{Calculations of energy levels}\label{subsec:levels}

\begin{figure}
\epsfig{file=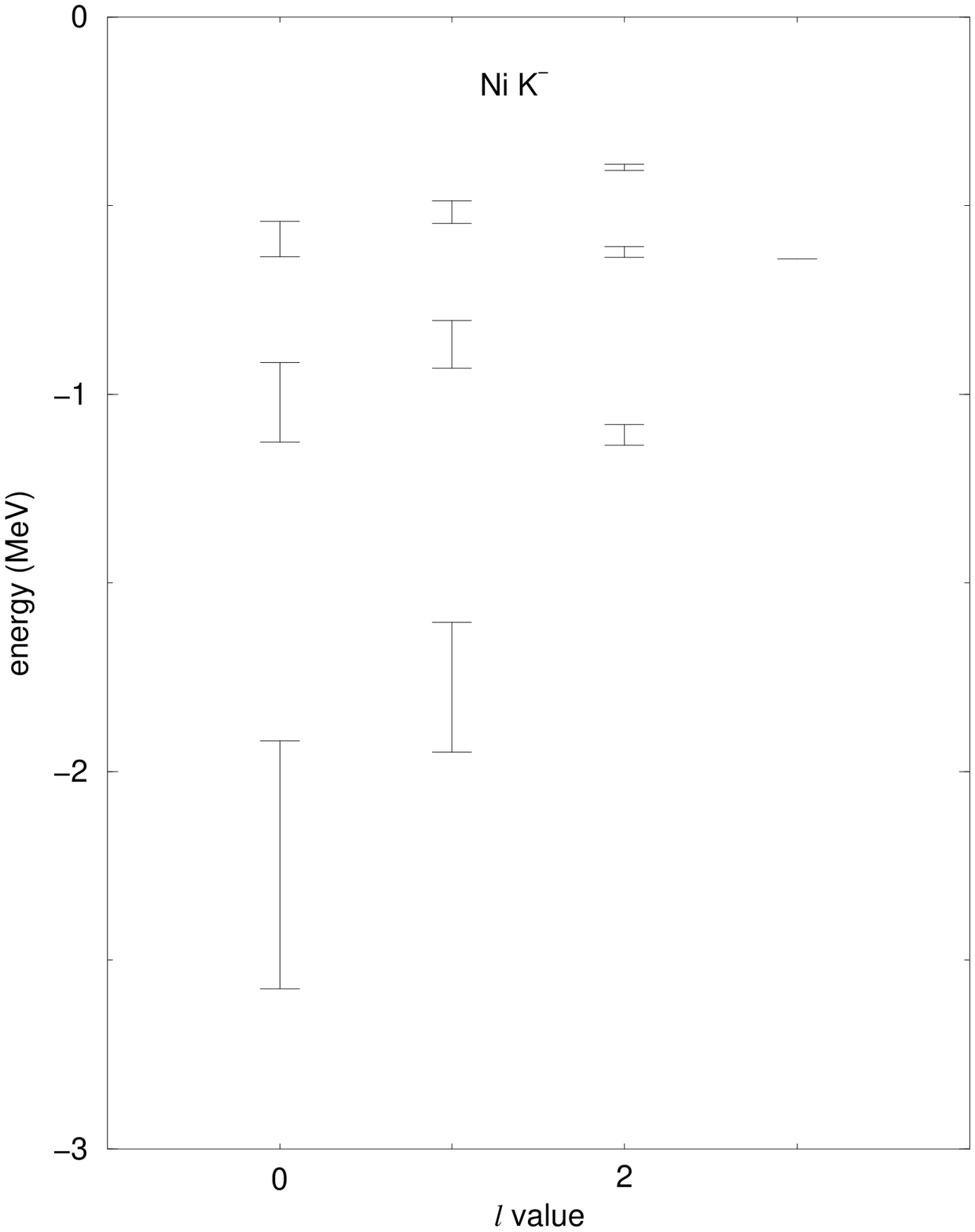,height=160mm,width=135mm,
bbllx=53,bblly=92,bburx=509,bbury=670}
\caption{Calculated energy levels for kaonic atoms of Ni
using the $t \rho$ potential specified in the text.
The  bars stand for the full width $\Gamma$ (=2 Im $B$) of the levels 
and the centers of the bars correspond to the energy ($-$ Re $B$).}
\label{fig:Nispect}
\end{figure}

\begin{figure}
\epsfig{file=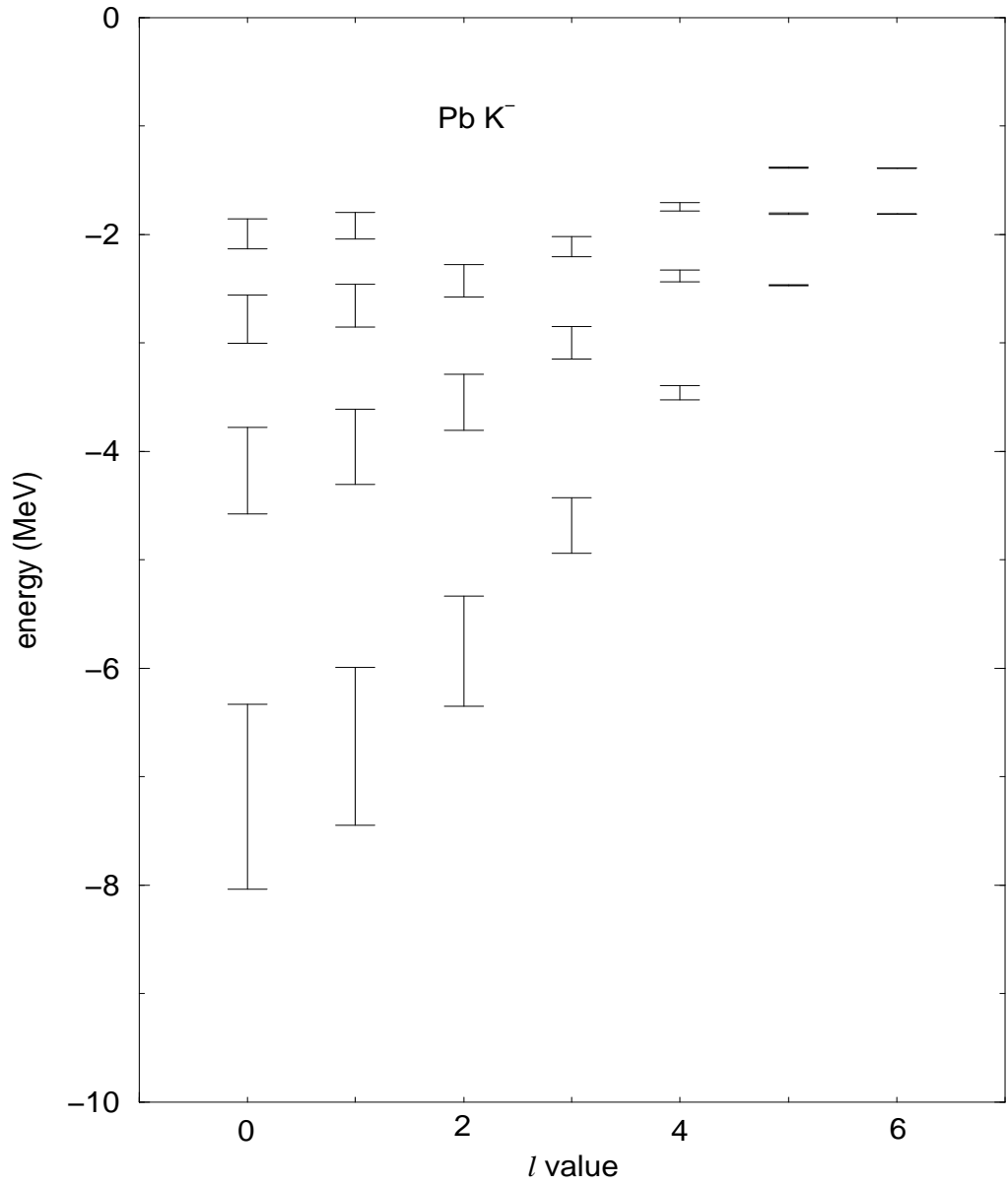,height=160mm,width=135mm,
bbllx=43,bblly=92,bburx=509,bbury=670}
\caption{Calculated energy levels for kaonic atoms of Pb.
See caption of FIG. 1 for details.}
\label{fig:PbspectK}
\end{figure}

Calculations of  energies and widths of kaonic atom levels were made for
three representative nuclei over the periodic table:
carbon, nickel and lead using  different potentials as outlined above.
The results for carbon, shown in Ref. \cite{FGa99} will not be repeated
here.
Figure \ref{fig:Nispect} shows calculated energy levels for Ni for
several values of $l$. The bars stand for the full width $\Gamma$ of the
level and the centers of the bars correspond to the 
binding energies Re $B$.
The lowest lying level for each value of $l$ corresponds to a circular
orbit with radial number $n=l+1$. The higher levels are for increasing
values of $n$.
The 4$f$ level is the last one populated in the atomic cascade
process and its position and width are known experimentally.
Similar results for Pb are shown in Fig. \ref{fig:PbspectK}. In this example
the last experimentally observed level is 7$i$. 
It is seen that the energy levels
are quite well defined also in this example of a heavy nucleus, but
it is clear that only with an $l$-selective process 
there is a chance to observe such states experimentally.
The dependence on the model is
found to be negligibly small; replacing the $t\rho$
potential by the DD potential or replacing the MAC density by the SP
density change the widths by less than 5\% and change the
binding energies by typically 3\% of the width, thus making the calculated
spectra essentially model independent, {\it provided} one employs
an optical potential which produces good fits to normal kaonic atom
data.

\begin{figure}
\epsfig{file=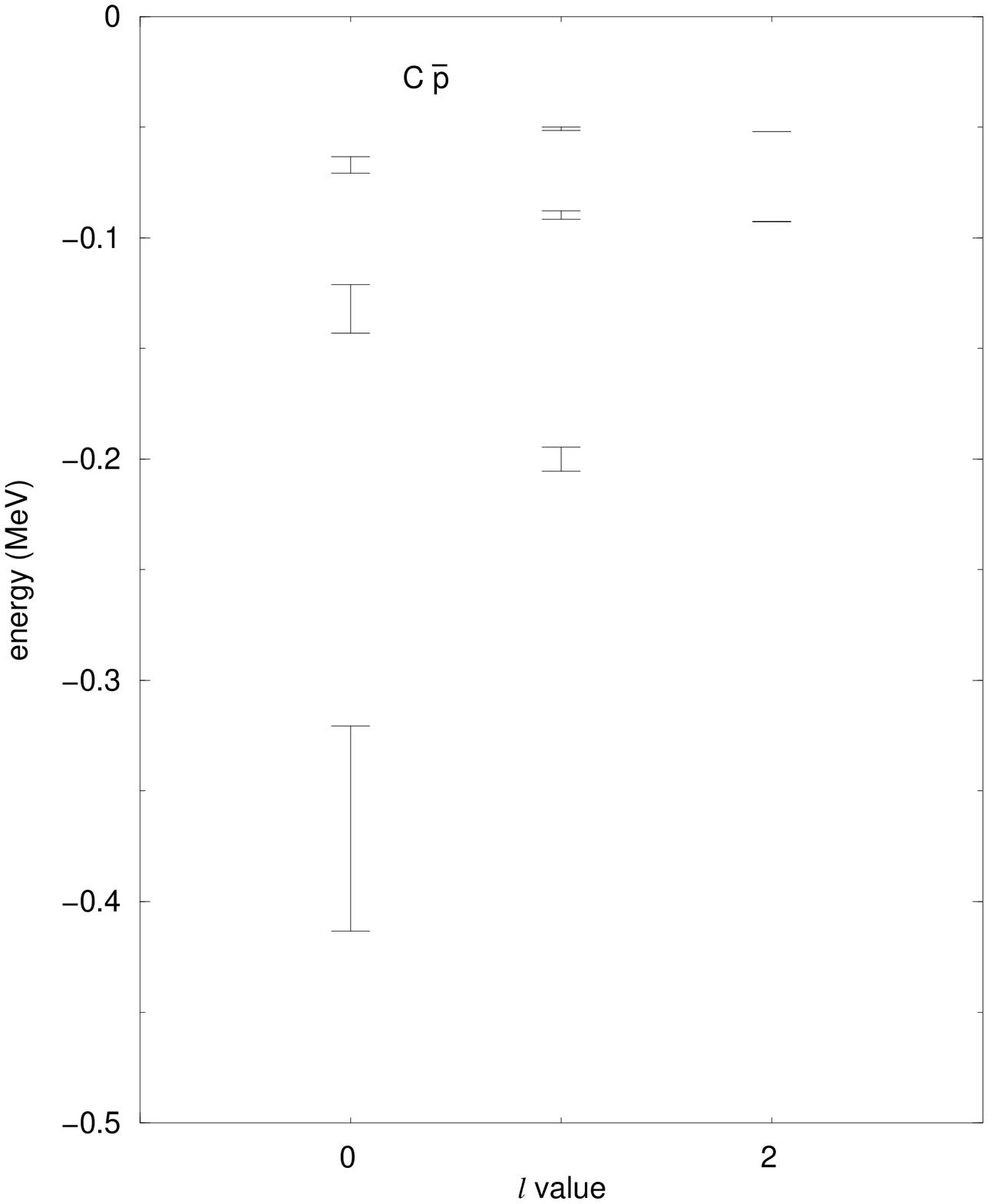,height=160mm,width=135mm,
bbllx=36,bblly=92,bburx=509,bbury=670}
\caption{Calculated energy levels for $\bar p$ atoms of carbon.
See caption of FIG. 1 for details.}
\label{fig:Cspect}
\end{figure}

\begin{figure}
\epsfig{file=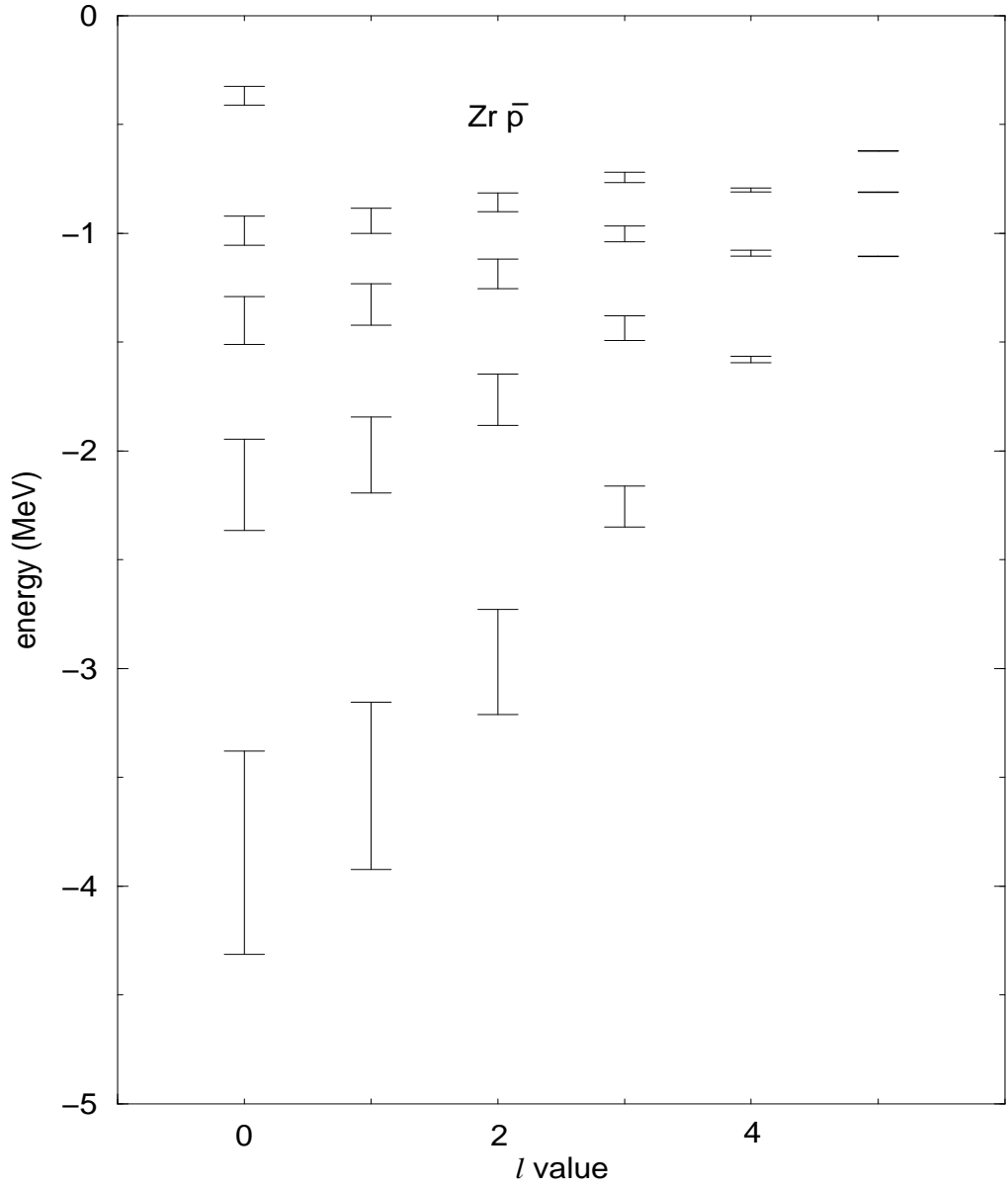,height=160mm,width=135mm,
bbllx=53,bblly=92,bburx=509,bbury=670}
\caption{Calculated energy levels for $\bar p$ atoms of zirconium.
See caption of FIG. 1 for details.}
\label{fig:Zrspect}
\end{figure}

\begin{figure}
\epsfig{file=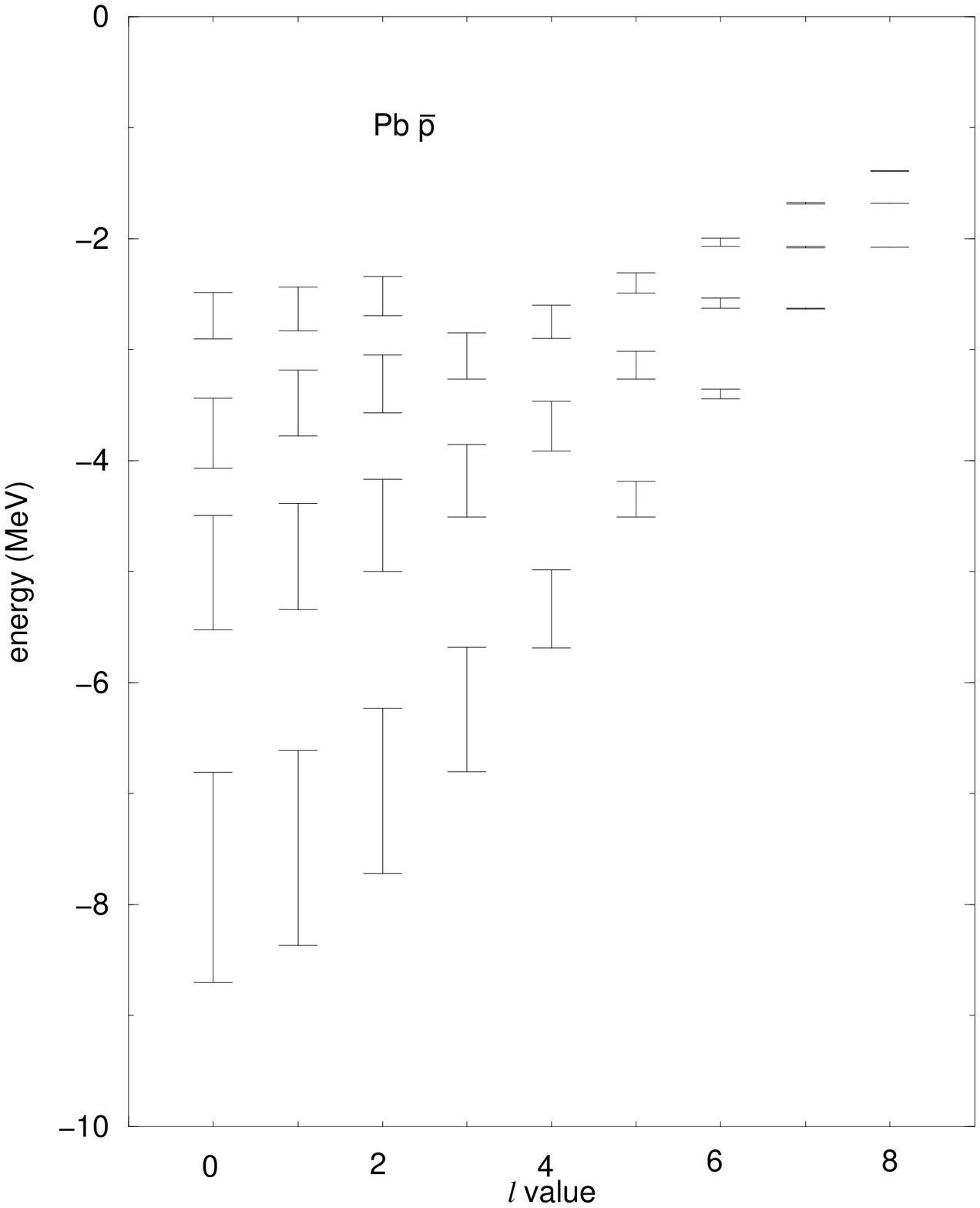,height=160mm,width=135mm,
bbllx=42,bblly=92,bburx=534,bbury=670}
\caption{Calculated energy levels for $\bar p$ atoms of lead.
See caption of FIG. 1 for details.}
\label{fig:Pbspectp}
\end{figure}

Calculations of binding energies and widths of levels of $\bar p$ atoms
were made with the $t\rho$ potential Eq. (\ref{equ:DD}) and with 
its $p$-wave extension Eq. (\ref{equ:EE1}) using SP
densities. Figures \ref{fig:Cspect} and \ref{fig:Zrspect} show calculated
energy level spectra for carbon and zirconium, 
respectively, elements for which normal
$\bar p$ atoms have been observed and the present potentials 
reproduce the data quite well.	
Figure \ref{fig:Pbspectp} shows predictions for
Pb, for which no experimental results are available.
For this heavy nucleus the levels are quite close to each other although
the spectrum is still well defined. 
Experimental observation of such levels clearly
must depend on selectivity in $l$ values. The dependence of the
results on the model is less than for deeply bound kaonic atoms, again
provided one uses potentials which fit data for normal states.

\subsection{Mechanism for narrowing of energy levels} \label{sec:narrowing}

\begin{figure}
\epsfig{file=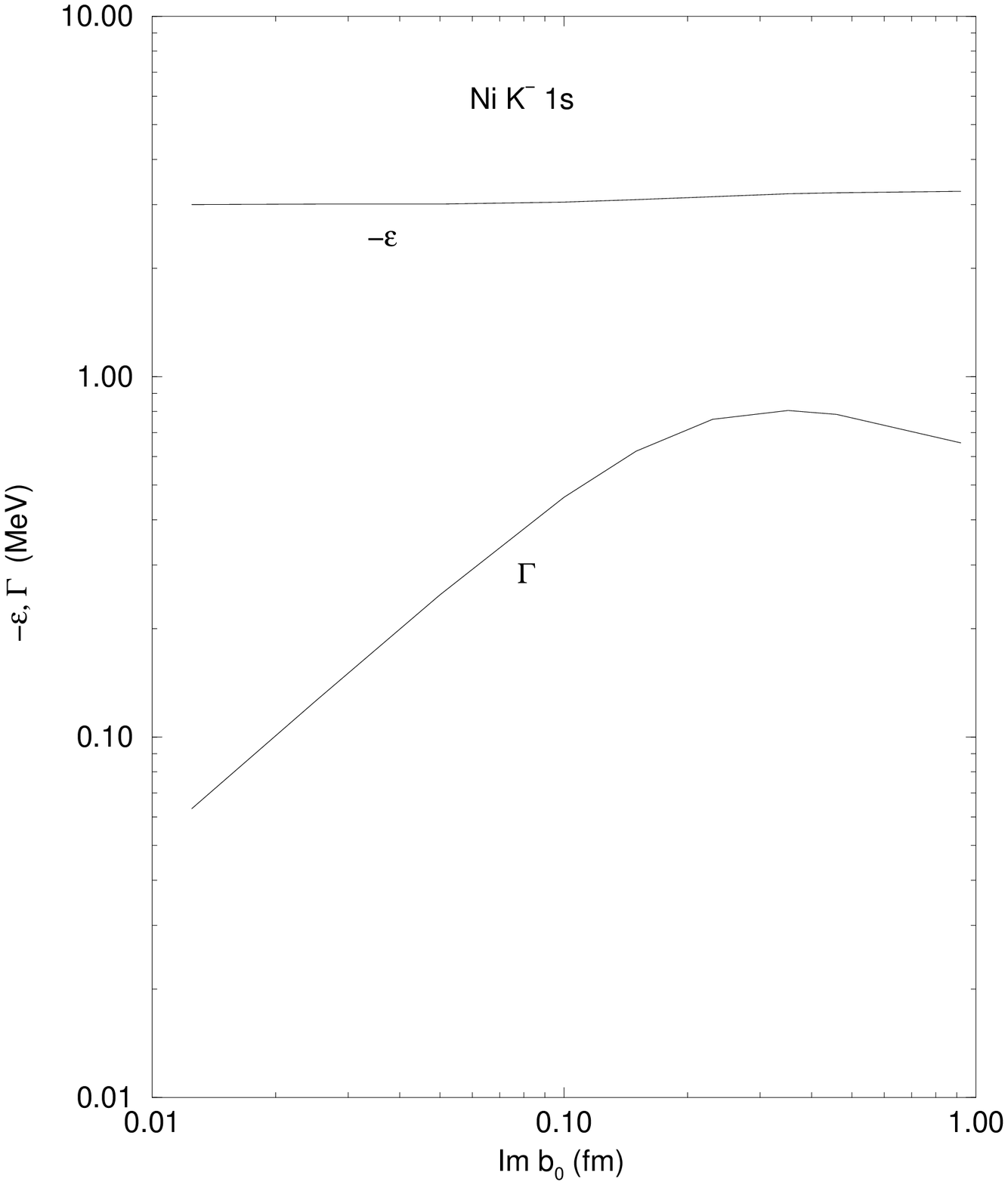,height=160mm,width=135mm,
bbllx=28,bblly=89,bburx=522,bbury=670}
\caption{Calculated strong interaction shifts 
($\epsilon$) and widths ($\Gamma$) as function of Im $b_0$ for
the 1$s$ level in kaonic Ni. Re~$b_0$=0.62~fm.}
\label{fig:KNi1seg}
\end{figure}

\begin{figure}
\epsfig{file=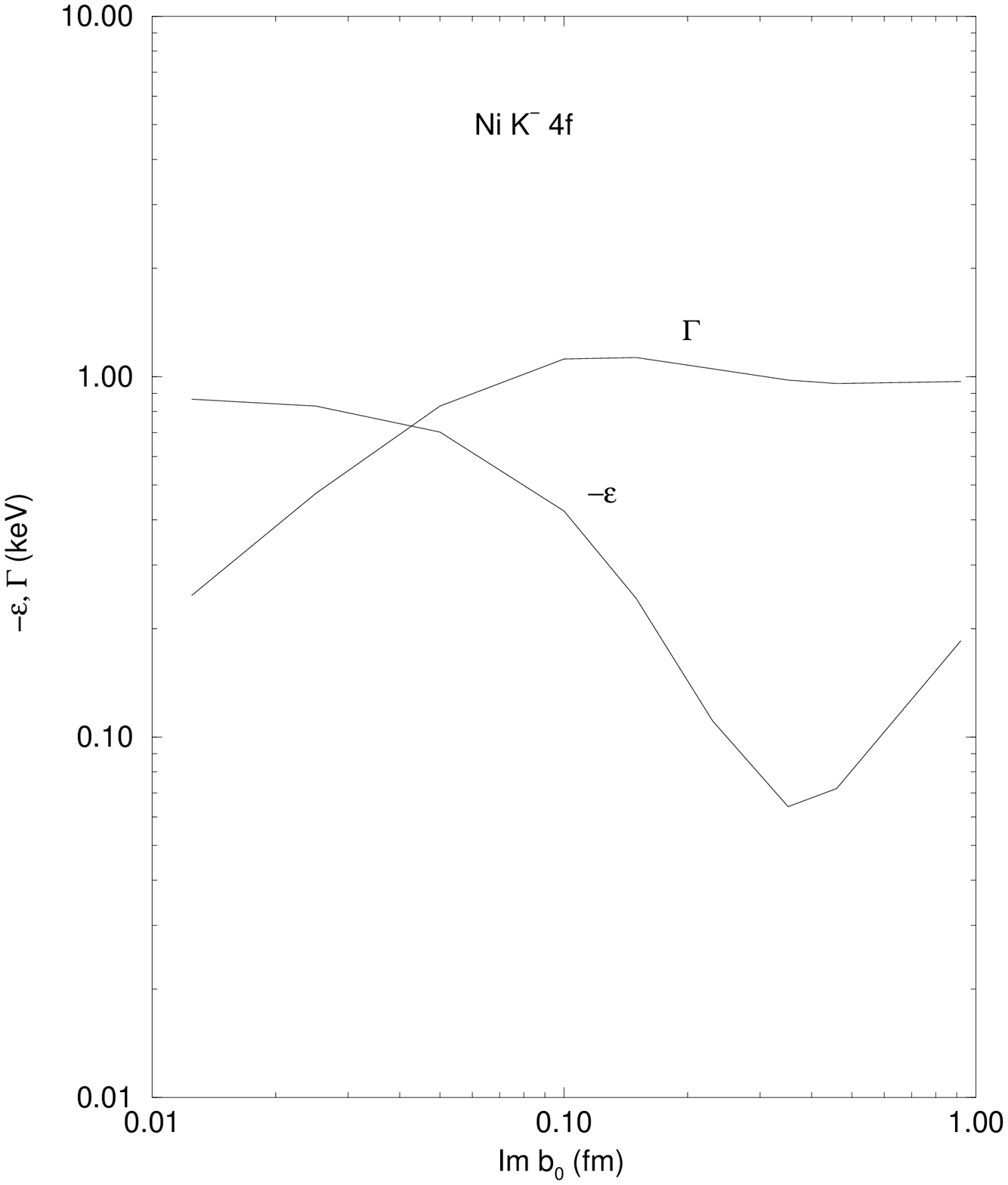,height=160mm,width=135mm,
bbllx=28,bblly=89,bburx=522,bbury=670}
\caption{Calculated strong interaction shifts and widths 
 as function of Im $b_0$ for
the 4$f$ level in kaonic Ni. Re~$b_0$=0.62~fm.}
\label{fig:KNi4feg}
\end{figure}

\begin{figure}
\epsfig{file=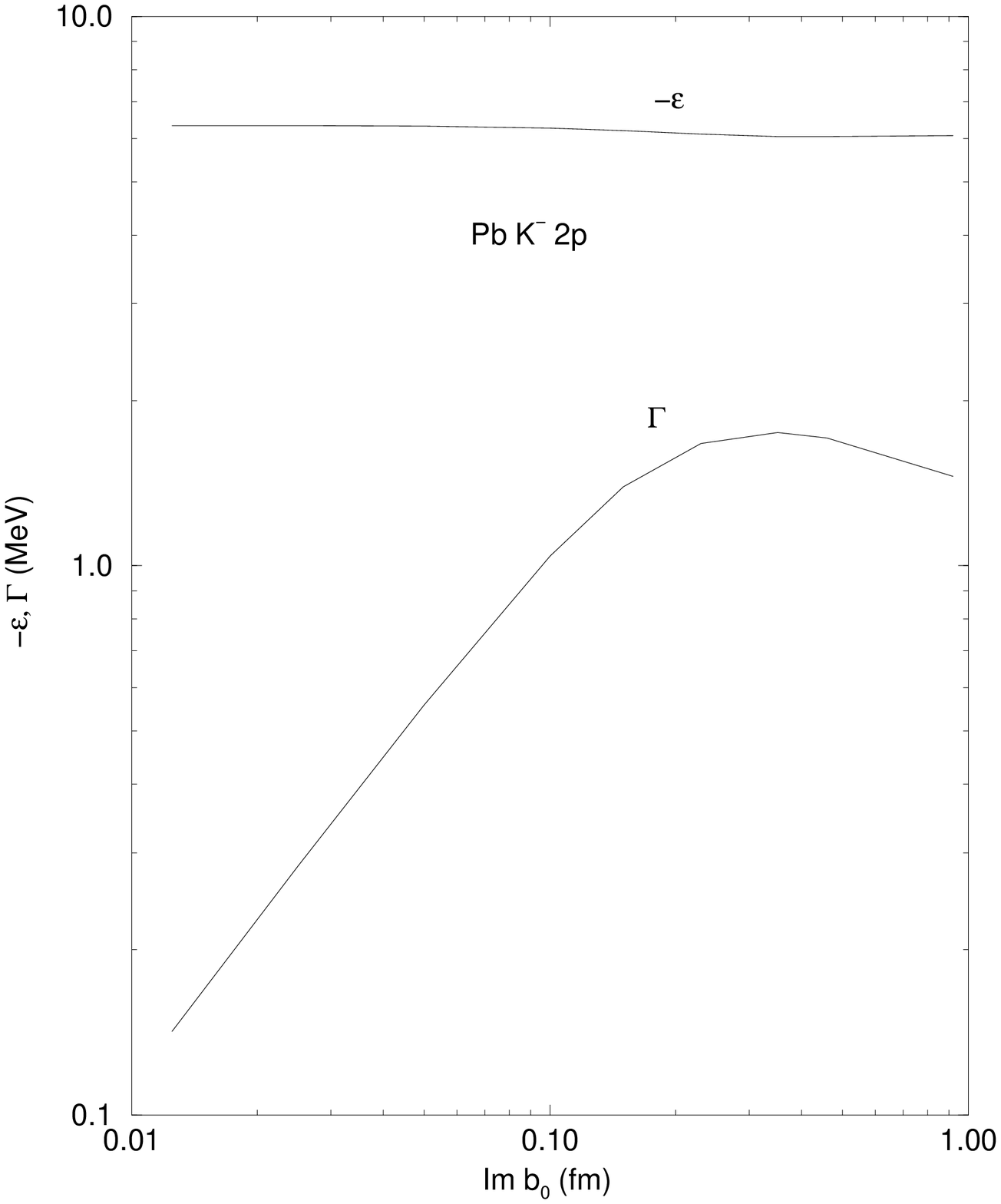,height=160mm,width=135mm,
bbllx=39,bblly=89,bburx=522,bbury=670}
\caption{Calculated strong interaction shifts and widths 
as function of Im $b_0$ for
the 2$p$ level in kaonic Pb. Re~$b_0$=0.62~fm.}
\label{fig:KPb2peg}
\end{figure}

The calculated narrow deeply bound levels in hadronic atoms demonstrated above
may seem surprising at first, particularly as they are obtained in
the presence of strongly {\it attractive} real potentials. This
phenomenon appears to be quite universal and the mechanism behind
it may be understood by studying the dependence of strong interaction
level shifts and widths on the hadron-nucleus potential
and by examining the wave functions.
Figure \ref{fig:KNi1seg} shows calculated shifts and widths for the 1$s$
level in kaonic Ni as functions of the imaginary part of $b_0$
when the real part of $b_0$ is being held at its nominal 
value, using the $t\rho$ potential.
It is seen that at about 15\% of the nominal value of Im $b_0$ (of 0.92 fm)
the width already saturates and then it goes slowly down, while the
shift stays essentially constant at a very large repulsive value.
The saturation of widths as function of the imaginary potential
is not confined to deeply bound states.
Figure \ref{fig:KNi4feg} shows similar results for the experimentally
observed 4$f$ level in Ni. 
The width saturates at 10\% of the nominal value
of Im $b_0$ but the (very small) shift changes rapidly so that for
Im~$b_0$=0.92 fm both shift and width agree with experiment. 
Figure \ref{fig:KPb2peg} shows similar behaviour for the 
2$p$ level in kaonic Pb.
These results may be understood by studying  the various
wave functions.

\begin{figure}
\epsfig{file=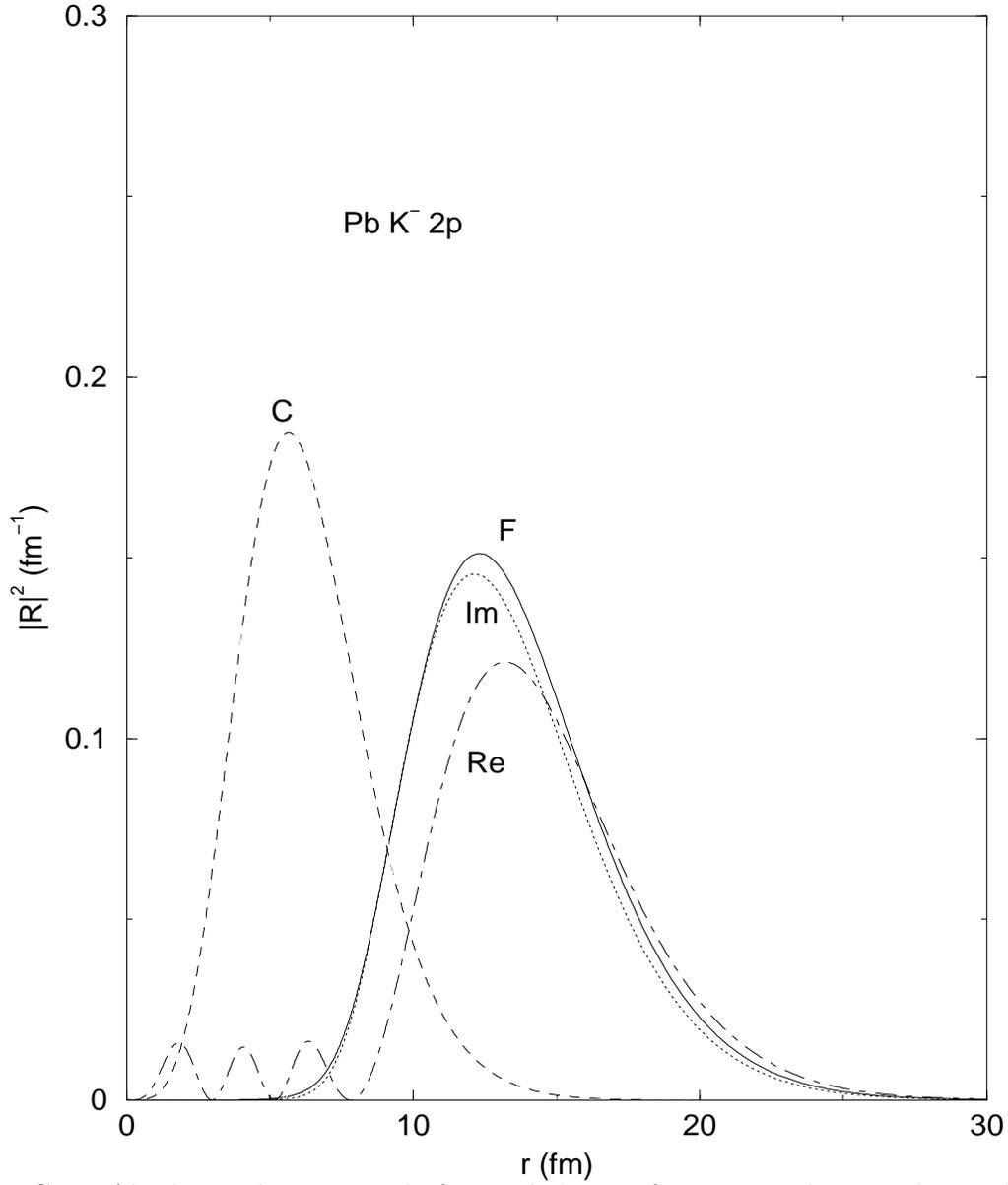,height=160mm,width=135mm,
bbllx=47,bblly=92,bburx=516,bbury=670}
\caption{Absolute values squared of 2$p$ radial wave functions in
kaonic Pb.  Dashed curve (C) for the Coulomb potential only,
solid curve (F) with the full optical potential added, dotted
curve (Im) for only the imaginary part of the potential added, 
dot-dashed curve (Re) for only the real potential added.}
\label{fig:KPb2pwf2}
\end{figure}

Figure \ref{fig:KPb2pwf2} shows the absolute value squared of the
radial wave function for the 2$p$ state in kaonic Pb for several combinations
of potentials. The dashed curve (C) is for the Coulomb interaction
only, which is always included.
Note that if this wave function is used in the integral Eq. (\ref{equ:gamma})
 a width of more that 100 MeV is obtained.
The solid curve (F) is for the full optical potential
added and it shows essentially total expulsion of the wave function from
the nucleus (whose rms radius is about 5.5 fm). The dotted curve (Im)
is when only the imaginary potential is added, and it shows that
the imaginary potential is dominant in determining the wave function. 
These two curves show that the
imaginary potenial  causes sufficient repulsion such that the overlap
between the atomic wave function and the nucleus becomes very small
compared to the pure Coulomb case, thus reducing dramatically the width
of the level, as is expected from Eq. (\ref{equ:gamma}).
The dot-dashed curve (Re) shows the wave function when only the real
optical potential is added. This {\it strongly attractive} potential
is seen to cause a substantial {\it repulsion} of the wave function
which is a phenomenon peculiar to hadronic atoms.
The three small inner peaks  preceding the main peak well outside the
nucleus indicate that three 
strongly bound states exist in this real potential, and since the real 
atomic wave function is 
approximately orthogonal to the wave functions of these states,
it develops nodes, giving rise to
the inner peak structure. This structure causes the main
{\it atomic} peak of the wave function to shift to larger radii 
compared to the Coulomb wave function, thus resulting in a repulsive shift.
The repulsion by the attractive real potential is by no means a universal
phenomenon. The effect depends on the $l$-value of the state and on the
positions of the internal nodes of the radial wave function; attractive
shifts are also possible although in most cases we have observed
repulson.
In the presence of the strong imaginary potential, the internal
structure gets suppressed but practically the 
same node structure remains for Re $\psi$.
The node structure inside the nucleus depends critically on the 
real optical potential and so are the energies of the strongly bound states.
Replacing the $t\rho$ potential for kaonic atoms by the DD potential,
the number of nodes changes but the position of the {\it outermost} 
node changes 
very little. The calculated energies of the atomic states hardly
change at all.

\section{Production of deeply bound atomic states}
\label{sec:prod}

The deeply bound atomic states discussed in the present work cannot be
observed by studying the spectra of X-rays emitted during the atomic cascade
process. Direct nuclear reactions, designed to implant strangeness for
$K^{-}$ atomic states, or antimatter for $\bar p$ atomic states, are
necessary in order to  produce such states. The lesson gained recently
from searching for and discovering deeply bound $\pi^{-}$ atomic
states \cite{Yam96,Yam98} is that, 
for a worthy candidate reaction, the
hadron has to be produced with as little momentum as possible, in
order to maximize the overlap with typical atomic wave functions.
Judged from the point of view of momentum transferred from the
projectile to the ejectile, this means looking for as low momentum
transfer reaction as possible,
say, $q \le $50 MeV/c. For pionic atoms the reaction chosen
was ($d,^3$He), using 600 MeV deuterons and looking for peaks
in the energy spectrum of the forward moving $^3$He nuclei at
energies corresponding to just below the $\pi^{-}$ production
threshold \cite{HTY91}. Further advantages of such production reactions are
that the suppressive effects of the nuclear absorption on the
production cross sections are weaker than for other reactions, and
that since the momentum transfer is minimal, so is the angular
momentum transfer: $\Delta l \approx 0$. This latter condition
ensures {\it selectivity}, namely that the atomic states most
copiously produced are those for which $l_{h}=l_{N}$, where $l_{h}$
is the orbital angular momentum of the produced atomic hadron and
$l_{N}$ is that of the target pick-up nucleon. Below, we discuss
several suitable reactions for producing $\bar p$ and $K^{-}$ deeply
bound states.

\subsection{The forward $(\bar p, p)$ reaction}

In this reaction an incoming antiproton hits a proton in the target
nucleus $^{A}Z$ and gets captured in $\bar p$ atomic states of the
resulting $^{A - 1}(Z - 1)$ core nucleus, while the proton flies off
in the forward direction. Neglecting binding effects, the momentum
transfer is zero in the forward direction, corresponding to
180$^0$ backward elastic scattering of antiprotons off protons
in the c.m. system. The $(\bar p, p)$ reaction was used at the LEAR
facility, CERN, at incoming momentum $p_{L} = 550$ MeV/c on C, Cu and
Bi targets, to search for deeply bound {\it nuclear} 
states \cite{Gar85}. The
measured energy spectra showed no clear evidence of a peak which could
be identified as a $\bar p$ bound state. However, it is far from clear
that the protons from this reaction on nuclei are observable at all on
top of a large physical background arising from $\bar p$ annihilation,
followed by secondary proton emission due to pion absorption. Baltz et
al. \cite{BDS85} calculated a `signal to background' ratio $R \sim 0.01-0.04$
in $^{16}$O for the leading $\Delta l = 0$ contribution, which would
be very difficult if not impossible to detect experimentally. In this
example, the more favorable ratio $R$ belongs to the $2p$ {\it
atomic} state, even though its forward excitation cross section is
suppressed by 3 orders of magnitude from about 0.5 mb/sr for the most
bound and very broad $\bar p$ nuclear $p$ state to about 0.5
$\mu$b/sr for this most bound, but narrow $(\Gamma \sim 28$ keV)
$\bar p$ atomic $2p$ state. When the relatively large cross section
of nuclear states is smeared in accordance with the large width, it
yields even poorer signal than for narrow atomic states. However, the
ratio $R$ introduced in that work implicitly assumes perfect energy
resolution, whereas the actual resolution $\Delta E \sim 1$ MeV in the
experiment \cite{Gar85} is considerably larger than the natural line width of
the $\bar p$ atomic $p$ states, so the ratio $R$ would have to be
reduced by roughly a factor 30.

Gibbs and Kaufmann \cite{GK84} observed that in certain favorable cases the
$(\bar p, p)$ cross section to atomic states are enhanced due to a
coincidence of a relatively sizable radial overlap between the bound
proton and antiproton and a large spectroscopic factor for this proton
orbit in the target ground state. In particular, they evaluated the
transition from the $2s$ proton state in $^{31}$P to the $1s~\bar p$
atomic state in the core nucleus of $^{30}$Si which in their
calculation was bound by 1.15 MeV with a width $\Gamma = 106$ keV.
The forward $(\bar p, p)$ cross section, as revised in Ref. \cite{BDS85}, is
about 10 $\mu$b/sr, over an order of magnitude larger than for the
example of $^{16}$O target discussed above. Since the $2p~\bar p$
atomic state in $^{30}$Si should lie only about 250 keV higher than
the $1s$ atomic state, it is clear that an extremely high
resolution $(\Delta E \sim 100$ keV) is absolutely essential in
searching for the few enhanced atomic $\bar p$ states. The forward
$(\bar p, p)$ cross section for the production of the $2p$ atomic
state $(\Delta l = 1)$ is smaller by about a factor 17 than for
producing the $1s$ atomic state $(\Delta l = 0)$ so it may be safely
neglected at $\theta = 0^0$. However, the $\Delta l = 1$
production cross section of the $2p$ atomic state peaks at about
6$^0$, where it assumes a comparable value to that of the
$\Delta l = 0$ cross section for the $1s$ atomic state.

For heavier $\bar p$ atoms, the lowest lying $1s$ and 
$2p$ atomic states lie
practically on top of each other, judging from Figs. (4, 5) for the
$\bar p$ atomic spectra in Zr and Pb. The $2p$ atomic state could be
explored by doing the $(\bar p, p)$ reaction at the very forward
direction on the $2 p_{1/2}$ proton in $^{89}$Y or $^{90}$Zr. A rough
estimate of the forward cross section gives 0.1 $\mu$b/sr. The $1s$
atomic state could be explored by doing the $(\bar p, p)$ reaction at
the very forward direction on the $3s_{1/2}$ protons in $^{208}$Pb.
The  isomeric ${11/2}^{-}$ state of the daughter core
nucleus $^{207}$Tl at excitation energy 1.35 MeV could be used to
explore the $\Delta l = 0$ component of the $(\bar p, p)$ reaction in
$^{208}$Pb, due to the filled $h_{11/2}$ proton shell, leading to the
$6h~\bar p$ atomic state. As seen in Fig. 5, the only states which
overlap this state would require $\Delta l \geq 2$ for their
excitation, which might then be neglected at the very forward direction.

\subsection{Production of $K^{-}$ atomic states}

This subject has not been considered quantitatively, to the best of
our knowledge. 
Hirenzaki et al. \cite{HOT99} very recently proposed to use the $(K^{-},
\gamma)$ reaction in order to search for deeply bound $K^{-}$ atomic
states. We note that the momentum transfer in this reaction is less
favorable for this purpose than that of the $(K^{-},p)$ reaction
discussed below. For example, for slow $K^{-}$ mesons resulting
from the decay of $\phi(1020)$  at rest, the momentum transfer in
the $(K^{-}, \gamma)$ reaction is about $q \sim 110$ MeV/c, more than
twice as large as for the $(K^{-}, p)$ reaction. It is also about
twice as large as the momentum transfer, for 20 MeV incident pions,
in the $(\pi^{-}, \gamma)$ reaction which was tried without much
success at TRIUMF \cite{Ray97} to search for deeply bound $\pi ^{-}$
atomic states. The main difficulty which must be overcome in trying to
isolate a low energy photon (or hadron) signal is the considerable
low-energy background expected from competing absorption processes.

The ($K^{-},p$) reaction for producing $K^{-}$ atomic states appears 
conceptually the closest one to $(\bar p, p)$ for $\bar p$ atomic 
states. Here, an incoming $K^{-}$ meson hits a target
proton and gets captured in $K^{-}$ atomic states of the daughter
$^{A-1}(Z-1)$ core nucleus, while the proton flies off in the forward
direction. However, since now the masses are unequal, $m_{K} \not=
m_{N}$, the momentum transfer in the ($K^{-}, p$) reaction differs
from zero except at rest. Neglecting binding energy corrections and
nuclear recoil, the momentum transfer in the forward direction is
given by:

\begin{equation}
	q\left( {0^0} \right)=\left[ {T_L\left( {T_L+2m_N} \right)}
	\right]^{{\raise3pt\hbox{$\scriptstyle 1$} \!\mathord{\left/
	{\vphantom {\scriptstyle {1 2}}}\right.\kern-\nulldelimiterspace}
	\!\lower3pt\hbox{$\scriptstyle 2$}}}-p_L \ \   ,
	\label{eq:q1}
\end{equation}
where $p_{L}$ and $T_{L}$ are the $K^{-}$ incoming momentum and
kinetic energy, respectively, in the laboratory (L) frame. For a
typical low-momentum $K^{-}$ beam at the Brookhaven AGS, with $p_{L} =
600$ MeV/c, the momentum transfer given by Eq. (\ref{eq:q1}) is 182
MeV/c, which is too high to be useful. Lowering the incoming momentum
further down to $p_{L} = 400$ MeV/c, the resulting value
$q(0^0) = 135$ MeV/c still appears too high.

A source of low-energy $K^{-}$ mesons is now available at the
Frascati $\Phi$-Factory $e^{+}e^{-}$ collider DA$\Phi$NE due to the
decay at rest of the $\phi (1020)$ meson to $K^{+} K^{-}$ pairs.
For the appropriate value of $p_{L} = 127$ MeV/c and $T_{L} = 16$
MeV, Eq. (\ref{eq:q1}) gives $q(0^0) = 47$ MeV/c which is
sufficiently small to make DA$\Phi$NE a possible site for searching for
deeply bound $K^{-}$ atomic states.  Although forward cross sections
as large as 0.1 $\mu$b/sr may be expected,
a major problem in doing this
experiment, as mentioned above,
 would be the need to detect unambiguously
the low energy outgoing proton,
if one is interested in deeply bound atomic states which are inaccessible to
the cascade process. (Otherwise X-rays can be detected from such
stopping $K^{-}$). If the decay at rest of the $\phi (1020)$ meson were
to take place at close proximity to a nucleus such that the kaonic
atom recoiled as a whole, then the signature of such a reaction would be
a peak in the energy spectrum of the associated $K^+$ above the end-point
of the normal $K^+$ spectrum. However, the relatively high momentum
transfer in this process, $q=$ 181 MeV/c, does not make it a favorable
candidate reaction.

Another idea due to members of the GSI collaboration searching for
meson-nuclear bound states \cite{GKY99} is to form the $\phi$ meson in the
nucleus using a quasi-free reaction, e.g.

\begin{equation}
	\bar p + p \to \pi^0 + \phi \ \ ,
	\label{eq:ann1}
\end{equation}
or similarly
\begin{equation}
	\bar p + n \to \pi^{-} + \phi \ \ ,
	\label{eq:ann2}
\end{equation}
or even
\begin{equation}
	\bar p + p \to \omega^0 + \phi \ \ ,
	\label{eq:ann3}
\end{equation}
where the lighter meson in the final state is observed in the forward
direction. For $\bar p$ incoming momentum $p_{L}$ = 5.3 GeV/c, the
$\phi$ meson is produced at rest in the reactions
(\ref{eq:ann1},\ref{eq:ann2}). The reaction (\ref{eq:ann3}) enables
the use of lower energy antiprotons, with $p_{L} = 1.4$ GeV/c, to
get $\phi$ at rest. The $\phi$ mesons thus formed decay into
$K^{-}K^{+}$ pairs, providing a source of low-energy $K^{-}$ `beam'
tagged by the observation of $K^{+}$ in the opposite direction.

It would be interesting to look for secondary nuclear interactions 
of the $\phi$ meson prior to its decay. A suitable
reaction in this respect is the
recoilless $(\phi, K^{+})$ reaction:

\begin{equation}
	\phi + {^{A}Z} \to K^{+} + {^{A}_{K^{-}}Z} \ \ ,
	\label{eq:phi}
\end{equation}
where ${^{A}_{K^{-}}Z}$ stands for the kaonic atom made by attaching
a $K^{-}$ meson to the nuclear target ${^{A}Z}$. Neglecting nuclear
recoil, the momentum transferred in this reaction from the $\phi$
to the forward $K^{+}$ meson is zero provided

\begin{equation}
	p_{L} = m_{\phi}((m_{\phi}/2m_{K})^{2}-1)^{1/2} = 262~~ {\rm MeV/c} .
	\label{eq:pl}
\end{equation}
To reach this incoming momentum, the reaction (\ref{eq:ann1}) could be
used at a considerably lower energy than argued above, namely at
$p_{L} = 0.95$ GeV/c, where the $\phi$ meson moves backward with a
momentum given by (\ref{eq:pl}). As for the reaction
(\ref{eq:ann3}) , it gives {\it at rest} $\phi$  mesons of about this
momentum. Furthermore, the photoproduction reaction

\begin{equation}
	\gamma + p \to p + \phi
	\label{eq:photo}
\end{equation}
also gives $\phi$ mesons with this backward momentum, for $p_{L}$ = 3.14
GeV/c.

The $(\phi, K^{+})$ reaction in some respects is analogous to the $(n,
p)$ reaction which was the one proposed originally by Toki and
Yamazaki \cite{TYa88} for producing deeply bound $\pi^{-}$ atomic
states. In both reactions the projectile ($\phi$ or $n$,
respectively) decays off shell

\begin{equation}
	\phi \to K^{+} + K^{-} \quad , \qquad n \to p + \pi ^{-} \quad ,
	\label{eq:decay}
\end{equation}
into the ejectile ($K^{+}$ or $p$, respectively) plus the negatively
charged hadron ($K^{-}$ or $\pi^{-}$, respectively) which gets
attached to the nuclear target in the atomic state described by a
coordinate-space wave function $\psi_{nlm}({\bf r})$. The
conditional probability to form such an atomic state is given by $|
\tilde \psi _{nlm}(\mbox{\boldmath $q$})|^{2}$, where $\tilde \psi$ is
the Fourier transform of $\psi$, and $\mbox{\boldmath $q$}$ is the
momentum transfer in the nuclear reaction. The cross section for
the $(\phi, K^{+})$ reaction Eq. (\ref{eq:phi}) 
may be estimated fairly reliably
in the plane-wave approximation (PWA), since the nuclear interactions
of both $\phi$ and $K^{+}$ are known to be rather weak. This is not
the case for the $(n, p)$ reaction, however, where the distortion
effects on the incident neutron and outgoing proton reduce the PWA
cross sections by typically two orders of magnitude \cite{NOs90}.
Following Ref. \cite{TYa88}, we estimate in the PWA the
forward-angle $(\phi, K^{+})$ cross sections for $q \ll p_{L}$ by

\begin{equation}
{{d\sigma ^{\rm PWA}} \over {d\Omega }}\approx {{g_{\phi K\bar K}^2}
\over {96\pi ^2}}{{p_L^2} \over {m_K}}\sum\limits_m {\left|
{\tilde \psi _{nlm}\left( \mbox{\boldmath $q$} \right)} \right|^2}\   ,
	\label{eq:xsec1}
\end{equation}
where only the dominant $q$ dependence, due to the atomic
wave function $\tilde \psi$, was retained. The coupling constant
$g_{\phi K \bar K}$ is determined by the rate of the on-shell $\phi$
decay into $K^{+}K^{-}$ pairs. It also satisfies approximately the
F-type SU(3) symmetry relationship

\begin{equation}
g_{\phi K\bar K}^2={1 \over 2}g_{\rho \pi \pi }^2
	\label{eq:symm}
\end{equation}
where $g_{\rho \pi \pi} = 6.04$ is determined from the $\rho^{0}
\to \pi^{+}\pi^{-}$ decay.

For the `magic' value of $p_{L}$ (Eq. (\ref{eq:pl})),
$q \approx 2 p_{L} \sin (\theta/2)$; Eq. (\ref{eq:xsec1}) which is valid
only at small angles assumes then the form

\begin{equation}{{d\sigma ^{\rm PWA}} \over {d\Omega }}
\quad \simeq \quad 0.155 \times
(2l+1)\left| {\tilde R_{nl}(q)} \right|^2\ \quad {\rm mb}\    ,
	\label{eq:xsec2}
\end{equation}
where the momentum space radial wave function $\tilde R_{nl}$
(dimension  fm$^{3/2}$) is related to $R_{nl} = u_{nl}(r)/r$ in
coordinate space [c.f. the discussion centered 
about Eqs. (\ref{equ:ortho},\ref{equ:norm}) in
Sec. \ref{sec:meth}] by

\begin{equation}
\tilde R_{nl}\left( q \right)=\sqrt{4\pi}~ 
%^{{\raise3pt\hbox{$\scriptstyle 1$} \!\mathord{\left/ {\vphantom
%{\scriptstyle {1 2}}}\right.\kern-\nulldelimiterspace}
%\!\lower3pt\hbox{$\scriptstyle 2$}}}
{\rm i}^l\int\limits_0^\infty
{j_l\left( {qr} \right)R_{nl}\left( r \right)r^2dr}\   .
	\label{eq:Fourier}
\end{equation}
For small values of $q$, the wave functions $\tilde R_{l=0}(q)$
decrease rapidly with $q$, whereas the wave functions $\tilde R _{l
\not= 0}(q)$ start from zero at $q = 0$ and reach their respective
maxima at finite values $q = q_{\rm max}^{(l)}$; the larger $l$ is, the
larger $q_{\rm max}^{(l)}$ is.
Moving gradually away from $\theta = 0~(q = 0)$,
the $(\phi, K^{+})$ reaction will favor formation of $K^{-}$ atomic
states with increasingly larger $l$ values. We have evaluated this
momentum space wave function for $K^{-}$ atomic $1s$ states at $q = 0$
across the periodic table. For the $1s$ states in Ni and Pb, c.f. Figs.
\ref{fig:Nispect} and \ref{fig:PbspectK},  the values of $\tilde R _{1s}(0)$ are

\begin{equation}
	{\rm Ni} : \quad 265 - {\rm i} 49.5~{\rm fm}^{3/2} \quad ,
        \qquad {\rm Pb}:\quad 268 + {\rm i} 116~{\rm fm}^{3/2} \quad ,
	\label{eq:rad}
\end{equation}
showing little variation as function of $Z$. This is a direct
consequence of the strong repulsion induced by Im $V_{\rm opt}$ which
keeps $R_{nl}(r)$ apart in spite of the increasingly attractive
$V_{c}(r)$ as function of $Z$. Substituting these values in Eq.
(\ref{eq:xsec2}) we obtain

\begin{equation}
	{{d \sigma^{\rm PWA}}\over{d \Omega}} (0^{\circ}) \quad
        \simeq \quad 11.3~{\rm b}~({\rm Ni})
	\quad , \quad 13.2~{\rm b}~({\rm Pb}) \quad .
	\label{eq:barns1}
\end{equation}
Using distorted waves for the $\phi$ and $K^{+}$ is not expected to
reduce these values by more than a factor of 5, judging for example
from the distorted wave calculations reported by Dover and Gal
\cite{DGa83} for the production of surface $\Xi$ hypernuclear states
in the $(K^-,K^+)$ reaction. 
These fantastically
large cross sections should not come as a surprise. In this reaction
the incoming $\phi$ meson serves as a very efficient source of
$K^{-}$ mesons which are then scattered quasi-elastically into well
localized {\it atomic} regions of coordinate space. If one could
scatter elastically $K^{-}$ mesons off an absorptive object of size
$a \sim 12$ fm (about which the $1s$ $K^{-}$ atomic states in Ni 
and Pb are
localized), the {\it diffractive} forward elastic scattering
cross section would be

\begin{equation}
	{{d \sigma^{\rm el}}\over{d \Omega}} (0 ^{\circ})
        \sim ({{1}\over{2}} p_{L}a)^{2}a^{2}=91.4~{\rm b} \quad ,
	\label{eq:barns2}
\end{equation}
for $p_{L}= 262$ MeV/c [Eq. (\ref{eq:pl})].
Note the $a^4$ dependence on the size parameter $a$. 
Thus, the diffractive $(\phi,
K^{+})$ reaction to these $1s$ $K^{-}$ atomic states 
exhausts less than 15\% of the bound suggested by the elastic scattering.
For realistic situations, however, finite angular resolution will
effectively reduce the above diffractive cross sections by 1-2
orders of magnitude.

More quantitative work is needed to embed the $(\phi, K^{+})$ reaction
within a full reaction calculation which considers also the
propagation and decay of the $\phi$ meson in the nuclear medium
subsequent to its formation.

\section{Summary and conclusions}
\label{sec:summ}

Calculations of binding energies of $K^-$ and $\bar p$ atoms
based on {\it realistic} optical potentials, i.e. potentials that
lead to good agreement between measured and calculated strong
interaction effects in hadronic atoms, revealed rich spectra
of well separated energy levels in regions inaccessible
to the common X-rays cascade process.
As both types of hadronic atoms
are characterised by strongly attractive and absorptive potentials,
the results may seem surprising at first sight. The relative
narrowness of these deeply bound states is found to result from the
effective repulsion due to the strong absorption, which suppresses
substantially the overlap between atomic wave functions and the
nucleus, thus leading to greatly reduced widths. The effective
repulsion is dominated by the imaginary part of the potential,
whereas the effects of the real potentials are marginal.

The calculated spectra of deeply bound states show remarkable
insensitivity to the details of the potentials, provided one
uses potentials that fit hadronic atom data for `normal' states.
However, experimental observation of deeply bound states may
provide new insight into the interaction of the respective
hadrons with nuclei at rest,
because the overlap of the atomic wave functions of deeply
bound states with nuclei are significantly larger than the overlap
for normal states. In order to observe such deeply bound states
one must find processes where a $K^-$ meson or a $\bar p$ are formed
with a relatively small (50 MeV/c) momentum transfer in an
$l$-selective process. Several such examples were given in the
preceding section. Particular emphasis was placed on reactions
which use low energy $\phi$ mesons as a doorway to inject $K^-$
mesons into deeply bound $K^-$ atomic states.

\vspace{10mm}

This research was partially supported by the Israel Science Foundation.

\end{document}